\documentclass[12pt,a4paper]{article}
\usepackage[english]{babel}
\usepackage{hyperref}
\usepackage{amssymb}
\usepackage{epsfig}
\usepackage[cm]{fullpage}

\title{Study of the effect of cost policies in the convergence of selfish strategies in Pure Nash Equilibria in Congestion Games}

\author{Vissarion Fisikopoulos}

\begin{document}

\newtheorem{Lemma}{Lemma}
\newtheorem{scetch}{Proof Scetch}

\maketitle

\vspace{1cm}
\begin{center}
Extended Abstract\footnote{The complete thesis is available, in greek format only, here: \htmladdnormallink{http://users.uoa.gr/$\sim$vfisikop/projects/thesis.pdf}{http://users.uoa.gr/~vfisikop/projects/thesis.pdf}}
\end{center}

\abstract{In this work we study of competitive situations among users of a set of global resources. More precisely we study the effect of cost policies used by these resources in the convergence time to a pure Nash equilibrium. The work is divided in two parts. In the theoretical part we prove lower and upper bounds on the convergence time for various cost policies. We then implement all the models we study and provide some experimental results. These results follows the theoretical with one exception which is the most interesting among the experiments. In the case of coalitional users the theoretical upper bound is pseudo-polynomial to the number of users but the experimental results shows that the convergence time is polynomial.}

\section{Introduction}

\paragraph{}

General goal of the current work is the study of competitive situations among users of a set of global resources. In order to analyze and model these situations we use as tools, game theoretic elements \cite{OsbRub}, such as Nash Equilibria, congestion games and coordination mechanisms. Every global resource debit a cost value to its users. We assume that the users are selfish; ie. their sole objective is the maximization of their personal benefit. An \textbf{Nash equilibrium} (NE) is a situation in which no user can increase his personal benefit by changing only his or her own strategy unilaterally. 
 
\paragraph{}

More specific, we are interested in the KP-model or parallel links\footnote{Read \cite{parallel_link_game} for a survey on algorithmic problems in parallel links model.} model with $n$ users(jobs) and $m$ edges(machines) and we study convergence methods to pure Nash Equilibria, in which all the strategies a user can select are deterministic. Generally, a game has not always a pure Nash equilibrium. Although we are going to study cases in which there is always at least one Nash equilibrium. We define as \textbf{cost policy} of an edge the function which computes the cost of each user of this edge.

\section{Framework}

\paragraph{}
One method of convergence in a pure Nash equilibrium is, starting from an initial configuration, to allow all users to selfishly change their strategies (one after the other) until they reach a pure Nash equilibrium. We are interested in the convergence time to pure Nash Equilibria, that is the number of these selfish moves. Firstly, we study the makespan cost policy, in which each edge debits its total load to everyone that use it. In the most simple case, the whole procedure is divided into several steps. At each step, the \textbf{priority algorithm} choose one user from the set of users that benefit by changing their current strategy. For this model, named \textbf{ESS-model}, the convergence time is at the worst case exponential to the number of users. We present the effect of several priority algorithms to the convergence time and results for the major different cases of edges (identical, related, unrelated) \cite{ESS}. 
\paragraph{}
Another approach, with applications to distributed systems, is the concurrent change of strategies (\textbf{rerouting}) \cite{rerouting04} in which more than one user can change simultaneously his strategy. This model is more powerful than ESS because of its real life applications but we are not analyzing it in this work.
\paragraph{}
An extension to ESS-model is that of \textbf{coalitions}, in which the users can contract alliances. This model comes from cooperative game theory. In this case we have to deal with groups of users changing selfishly their group strategies. We restrict our attention to coalitions of at most 2 users in the identical machines introduced in \cite{coalitions}. The pairs of coalitional users can exchange their machines making a 2-flip move as a kind of pair migration. There is a pseudo-polynomial upper bound to the convergence time to NE in this model. 
\paragraph{}
Another model of convergence, a little different than the others stated above, is the construction of an algorithm that delegates strategies to the users unselfishly without increasing the social cost. Informally, social cost is a total metric of the system performance depending on the users strategies. This model is named \textbf{nashification} and the algorithm nashify provides convergence to a pure Nash equilibrium in polynomial number of steps without increasing the social cost \cite{nashification}.   
\paragraph{}
As far as the \textbf{coordination mechanisms} are concerned, they are a set of cost policies for the edges, that provides motives to the selfish users in order to converge to a pure Nash equilibrium with decreased social cost. 
\paragraph{}
In this paper, we study the effect of coordination mechanisms in the convergence time. We study the following cases:\\

\textbf{Cost Policies:} (1) Makespan, (2) Shortest Job/User First (SJF), (3) Longest Job/User First (LJF), (4) First In First Out (FIFO).\\

\textbf{Priority Algorithms:} (a) max weight job/user (maw), (b) min weight job/user (miw), (c) fifo, (d) random.\\

Note that each machine uses a cost policy and the next for migration user is chosen using a priority algorithm and the above combinations can result in \textit{linear}, \textit{polynomial} or \textit{exponential} convergence.

In the concept of coalitions there is a need of tight braking when we have to choose among pairs of users. We study two algorithms:\\  

\textbf{Coalition Priority Algorithms:} (i) max weight pair (map), (ii) min weight pair (mip).\\

Note that there is a arrangement of pairs comparing the subtraction of weights for each pair.

\section{Results}

The paper's results are divided in two categories: theoretical and experimental. 

\paragraph{Theoretical Results.} We study the convergence time for SJF, LJF and FIFO policies. Especially for FIFO we prove in identical machines case a tight linear bound and a pseudo-polynomial bound in unrelated machines case.

\begin{Lemma} \label{l1}
Under the FIFO cost policy in the ESS-model, there is an upper bound of $\frac{n^2}{2}w_{max}$ steps for convergence to NE in the unrelated machines case and $n-1$ steps in the identical machines case. This result is independent from the priority algorithm.
\end{Lemma}

\textit{Proof Sketch.} 
Note that in every step the cost of at least one user will be decreased so the cost of each user will not be increased. Also there is a potential function that decrease at least one unit at each step (under the integer weights assumption). Combining this with the max potential $P(0)\leq \frac{n^2}{2}w_{max}$ where $w_{max}$ the maximum weight we have the desired bound for unrelated machines. In the identical machines case we only have to observe that when a user migrates then it has no benefit to migrate again. This is true because at each step the minimum load did not increase so our result holds. $\square$

\begin{Lemma}  \label{l2} 
Under the LJF(SJF) cost policy in the ESS-model, the miw(maw) priority algorithm gives an upper bound of $n$ steps for the identical machines case.
\end{Lemma}

\textit{Proof Sketch.}
Similar with the FIFO policy we have to prove that when a user migrates then it has no benefit to migrate again. Let $N=\{1,2,\ldots,n\}$ the set of users sorted by not increasing weight order and $N_t=\{t+1,\ldots,n\}$ the set of users with weight greater or equal to that of user $t$. We prove the desired result by induction to the number of steps. $\square$
\paragraph{}
In \cite{coordination_selfish} there is a proof of a quadratic upper bound to the convergence time of unrelated machines using LJF and SJF policies but only for a restricted case. However, our experiments provide us with exponential lower bounds for the general case. 

\paragraph{}

\paragraph{Experimental Results.} We implement\footnote{The code in C is open and available here:\\ \htmladdnormallink{http://users.uoa.gr/$\sim$vfisikop/projects/thesis\_code/kp\_velt.c}{http://users.uoa.gr/~vfisikop/projects/thesis_code/kp_velt.c}} all the above mentioned models and analyze them experimentally. 

The main usage of this analysis is to provide us with lower bounds. In our experiments there are 3 parameters: the priority algorithm, the cost policy, and the priority algorithm of coalitions.
The following weight distributions are used:

\begin{list}{}{}
    \item {(a)}{} user's $10\%$ have weight $10^{n/10}$ and $90\%$ have weight $1$. 
    \item {(b)}{} user's $50\%$ have weight $10^{n/10}$ and $50\%$ have weight $1$. 
    \item {(c)}{} user's $90\%$ have weight $10^{n/10}$ and $10\%$ have weight $1$. 
    \item {(d)}{} random distribution from the set $[1,10^{n/10}]$.
    \item {(e)}{} each user has as weight his/her id ($w_1=1,w_2=2,\ldots,w_n=n$).
\end{list}

\begin{figure}[t]
  \hfill
  \begin{minipage}[t]{.45\textwidth}
    \begin{center}  
      \includegraphics[scale=0.3]{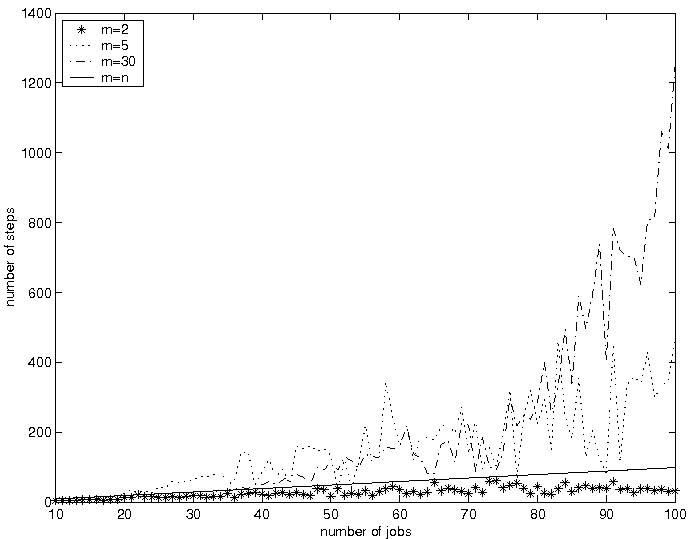}
      \caption{Maw algorithm with Makespan policy using mip algorithm for coalitions and weight case (d)}
      \label{fig:3}
    \end{center}
  \end{minipage}
  \hfill
  \begin{minipage}[t]{.45\textwidth}
    \begin{center}  
      \includegraphics[scale=0.3]{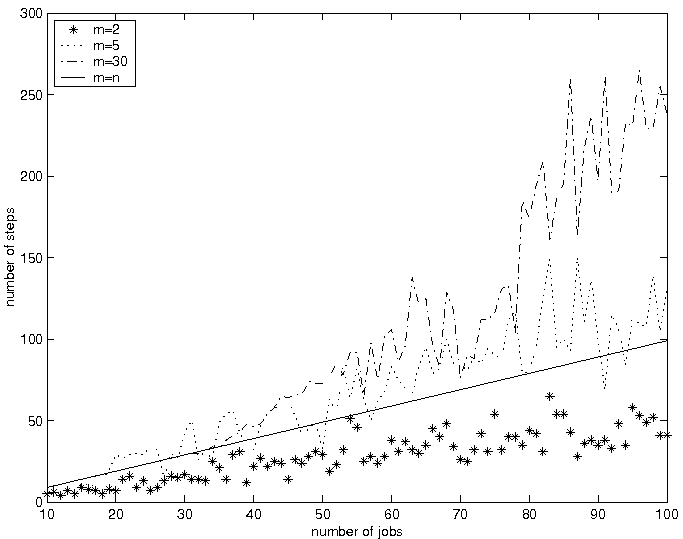}
      \caption{Maw algorithm with Makespan policy using map algorithm for coalitions and weight case (d)}
      \label{fig:4}
    \end{center}
  \end{minipage}
  \hfill
\end{figure}

 In all cases the experimental results follows the theoretical with one exception which is the most interesting among the experiments. In the case of coalitions with at most 2 users the theoretical upper bound is pseudo-polynomial to the number of users but the experimental results shows that the convergence time is polynomial. Especially the experiments has shown that the number of 2-flips is quadratic using mip algorithm and linear using map algorithm (Figures \ref{fig:3}, \ref{fig:4}). In addition to that, the 2-flips are a large percent of the total migration steps so we can conjecture that the pseudo-polynomial theoretical upper bound can improve to a polynomial one.

\begin{figure}[t]
  \hfill
  \begin{minipage}[t]{.45\textwidth}
    \begin{center}  
      \includegraphics[scale=0.4]{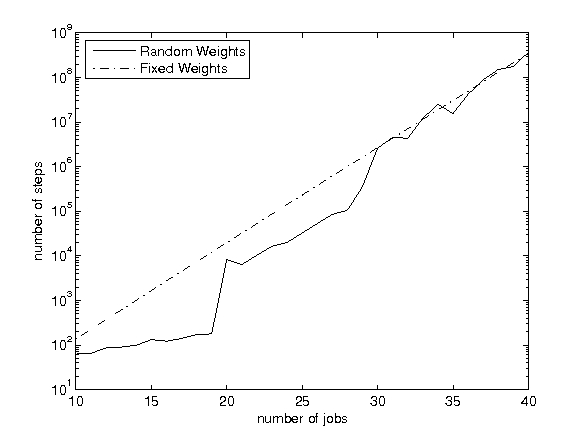}
      \caption{Maw algorithm with SJF policy and weight case (d) and (e) (y axis in logarithmic scale)}
      \label{fig:1}
    \end{center}
  \end{minipage}
  \hfill
  \begin{minipage}[t]{.45\textwidth}
    \begin{center}  
      \includegraphics[scale=0.4]{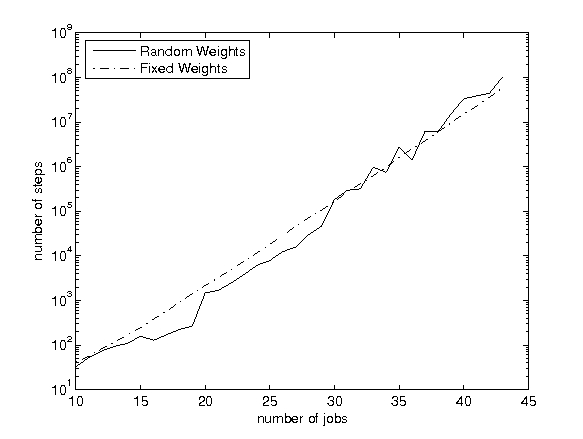}
      \caption{Miw algorithm with LJF policy and weight case (d) and (e) (y axis in logarithmic scale)}
      \label{fig:2}
    \end{center}
  \end{minipage}
  \hfill
\end{figure}

\section{Conclusions}

\paragraph{Identical Machines.} The linear convergence is achieved when the priority algorithm benefits from the cost policy and chooses sequences of migrating users in order to satisfy the property: ``\textit{a user is stabilized after his migration}''. In makespan and SJF policies this obtained by the maw priority algorithm. In LJF policy this obtained by the miw priority algorithm. In FIFO policy this is obtained by any priority algorithm. Additionally, FIFO has the advantage that each machine doesn't need the knowledge of its users' weights, which is useful in applications in which the users don't have the knowledge of their weights. 

The exponential convergence in identical machine case is achieved when the priority algorithm operates ``against'' the machines' policy. Intuitively, this happens because the algorithm chooses the most ``unstable'' user for migration. In SJF(LJF) policies this obtained by the maw(miw) priority algorithm. 

Table 1 sums up the results of convergence time to NE for the identical machine case. The lower bounds are constructed from the experiments (see for example figures \ref{fig:1} and \ref{fig:2}). The upper bounds for FIFO, SJF, LJF policies are lemmas \ref{l1} and \ref{l2} and the results for Makespan policy are taken from \cite{ESS}.

\begin{table}[t]\label{results1}
\centerline{
\begin{tabular}{|c||c|c|c|c|c|c|c|c|}
  \hline
  \textbf{\scriptsize IDENTICAL}
             & \multicolumn{2}{|c|}{\textbf{Makespan}}
             & \multicolumn{2}{|c|}{\textbf{FIFO}}
				 & \multicolumn{2}{|c|}{\textbf{SJF}}
				 & \multicolumn{2}{|c|}{\textbf{LJF}}
\\[1pt] \cline{2-3} \cline{3-4} \cline{4-5} \cline{5-6} \cline{6-7} \cline{7-8} \cline{8-9}
\textbf{\scriptsize MACHINES} 
	& Lower & Upper & Lower & Upper & Lower & Upper & Lower & Upper 
\\[1pt] \hline \hline	
\begin{tabular}{c}\qquad\\ \textbf{maw}\\\qquad \end{tabular}
	& $n$ & $n$ & $n$ & $n$ & $c^n$ &  & $n$ & $n$
\\[1pt] \hline 	
\begin{tabular}{c}\qquad\\ \textbf{miw}\\\qquad \end{tabular}
	& $n^2$ &  & $n$ & $n$ & $n$ & $n$ & $c^n$ & 
\\[1pt] \hline
\end{tabular}
}
\vskip-0.05cm
\caption{Lower and upper bounds of convergence time to NE for the identical machine case. The results for Makespan policy are taken from \cite{ESS}.}
\end{table}

\paragraph{Unrelated Machines.} In this case the upper bounds are either exponential or pseudo-polynomial and FIFO policy seems to be the best again. Table 2 sums up the results of convergence time to NE for the unrelated machine case.

\begin{table}[t]\label{results2}
\centerline{
\begin{tabular}{|c||c|c|}
  \hline
  \textbf{\scriptsize UNRELATED}
             & \textbf{Makespan}
             & \textbf{FIFO}
\\[1pt] \cline{2-3}  
\textbf{\scriptsize MACHINES} 
	& Upper & Upper  
\\[1pt] \hline \hline	
\begin{tabular}{c}\qquad\\ \textbf{maw}\\\qquad \end{tabular}
	& \begin{tabular}{c} $2mW_{max}$\\$+m2^{W_{max}/m+W_{max}}$ \end{tabular} & $\frac{n^2}{2}+W_{max}$  
   \\[1pt] \hline 	
\begin{tabular}{c}\qquad\\ \textbf{miw}\\\qquad \end{tabular}
	& $2^{W_{max}}$ & $\frac{n^2}{2}+W_{max}$  
\\[1pt] \hline
\end{tabular}
}
\vskip-0.05cm
\caption{Lower and upper bounds of convergence time to NE for the unrelated machine case. The results for Makespan policy are taken from \cite{ESS}.}
\end{table}

\section{Future Work} 

As a possible extension one can study other cost policies and priority algorithms and compare with the existing ones. The above models can be studied under the concept of \textbf{price of anarchy}. Is a good cost policy under the convergence time to NE concept also good under the price of anarchy concept? What is FIFO's price of anarchy? The cases of related and unrelated machines have also many open problems. Finally in games among coalitions it is interesting to study coalitions with more than 2 players. 

\section{Acknowledgements}

This work is a part of my bachelor's thesis at the University of Patras, Department of Computer Engineering \& Informatics.  It was supervised by Paul Spirakis and Spiros Kontogiannis.

\bibliographystyle{alpha}
\bibliography{bibliography}

\end{document}